# Lasing in an assembled array of silver nanocubes†

Mindaugas Juodėnas, [ID] *[a] Nadzeya Khinevich, [ID] [a] Gvidas Klyvis,[a] Joel Henzie, [ID] [b] Tomas Tamulevičius [ID] [a] and Sigitas Tamulevičius [ID] [a]

We demonstrate a surface lattice resonance (SLR)-based plasmonic nanolaser that leverages bulk production of colloidal nanoparticles and assembly on templates with single particle resolution. SLRs emerge from the hybridization of the plasmonic and photonic modes when nanoparticles are arranged into periodic arrays and this can provide feedback for stimulated emission. It has been shown that perfect arrays are not a strict prerequisite for producing lasing. Here, we propose using high-quality colloids instead. Silver colloidal nanocubes feature excellent plasmonic properties due to their single-crystal nature and low facet roughness. We use capillarity-assisted nanoparticle assembly to produce substrates featuring SLR and comprising single nanocubes. Combined with the laser dye pyrromethene-597, the nanocube array lases at 574 nm with <1.2 nm linewidth, <100 μJ cm$^{-2}$ lasing threshold, and produces a beam with <1 mrad divergence, despite less-than-perfect arrangement. Such plasmonic nanolasers can be produced on a large-scale and integrated in point-of-care diagnostics, photonic integrated circuits, and optical communications applications.

**New concepts**

Small plasmonic nanolasers based on surface lattice resonance are a unique counterpart to 2D distributed feedback lasers by exclusively utilizing localized surface plasmon resonance supporting scattering units. Although they have been extensively researched since 2013, these nanolasers have predominantly been fabricated using bottom-up techniques like physical vapor deposition, resulting in polycrystalline nanoparticles. This universally leads to diminished scattering properties compared to those calculated in numerical models, introducing additional absorptive losses due to grain boundary scattering of plasmons. Moreover, a critical question in the field has been: how much order is necessary for these plasmonic systems to sustain lasing modes? Interestingly, it has been found that perfect order is not essential; in fact, a certain degree of disorder can enhance lasing characteristics. Thus, our work demonstrates a conceptually new way of thinking about the fabrication of these systems: employing mass-produced monocrystalline silver nanocubes as building blocks and assembling them on templates for plasmonic nanolasers. Using template-assisted assembly with single-particle resolution on a large scale, our systems provide lasing feedback and coherent light emission on-par with polycrystalline counterparts in the existing research—but on a scalable platform. This insight challenges the conventional need for highly ordered, cleanroom-produced systems and presents a new method to create more cost- and time-efficient, robust plasmonic nanolasers.

## Introduction

The demand for miniaturization and improved efficiency of coherent light-emitting devices has steadily increased over time.[1] Small lasers operating at the nanoscale can benefit applications such as optical communications, sensing, bioimaging, photonic circuits, *etc.*,[2] where a superior size-to-efficiency ratio and decrease in power requirements are particularly attractive. Making nanolasers requires ultimate control of light fields at the nanoscale, and metasurfaces, 2D collections of nanostructures, are excellent at this task. In particular, plasmonic metasurfaces enable precise control over light–matter interactions leading to enhancements in light emission, confinement, and manipulation by squeezing electromagnetic fields to subwavelength scales in the near field.[3]

Plasmonic metasurfaces can feature surface lattice resonances (SLRs) – collective in-phase excitations formed by localized surface plasmons (LSP) of nanoparticles and diffractive photonic modes of the lattice. These SLRs have been extensively studied and reviewed.[4] One of the applications of these hybrid plasmonic–photonic modes is light amplification and stimulated emission.[1,5,6] When a SLR-supporting plasmonic system is in contact with a medium that can provide gain, the lattice resonance acts as a feedback mechanism, amplifies

[a] Institute of Materials Science, Kaunas University of Technology, K. Baršausko St. 59, Kaunas LT-51432, Lithuania. E-mail: mindaugas.juodenas@ktu.lt
[b] Research Center for Materials Nanoarchitectonics (MANA), National Institute for Materials Science (NIMS), 1-1 Namiki, Tsukuba, Ibaraki 305-0044, Japan
† Electronic supplementary information (ESI) available: More details on methods and materials and experiments. See DOI: https://doi.org/10.1039/d4nh00263f

photoluminescence (PL) and can generate light that is both spatially and temporally coherent.

The first SLR-based nanolasers were made using electron beam deposited gold[7] and silver[8] nanocylinders and organic dyes. Subsequent research expanded on this foundation to achieve wavelength tunability via the refractive index[9] and stretching of elastomer substrates,[10] direction of emission control via multiple lattice modes,[11–13] and switching using magnetic fields.[14] Apart from dyes, quantum dots[15–18] and up-converting nanoparticles[19] have also been used as gain materials. Emission wavelengths from UV using aluminum nanoparticles to NIR using gold have been demonstrated, and a combination thereof accomplished white-light lasing.[20]

While most reported SLR-based nanolasers are fabricated using either e-beam lithography[7,9,12,14,20–22] or the PEEL method (a combination of phase-shifting photolithography, etching, electron-beam deposition and lift-off of the film),[23] recent studies have shown that nanolasing can still occur even when a significant fraction of nanoparticles is absent from the array or the lattice order is decreased.[12,24,25] Interestingly, removing some fraction of particles can even help to outcouple light and increase the slope of the input–output curve while maintaining a similar threshold.[25] This intriguing finding suggests that defect-free lattice fabrication through lithography may not be strictly necessary and opens the door to bottom-up methods that can offer improved scalability and cost at the expense of accuracy and precision.

An attractive prospect in this regard is the use of colloidal nanoparticles.[26] Highly crystalline colloids can be synthesized on a large scale.[27] Crystallinity ensures superior scattering-to-absorbance ratios versus physically deposited polycrystalline materials with grain boundaries and rough facets that generate absorptive damping.[28,29] Furthermore, these colloids are available in a variety of shapes, which could enable intelligent engineering of hot spots and facilitate anisotropic Purcell enhancement of the gain media for tunable devices.[30,31] The challenge then lies in arranging these nanoparticles in a manner conducive to providing lasing feedback.

Template-assisted nanoparticle assembly is a robust technique that can address this task.[32–34] A variation of it has been used to produce a colloid-based nanolaser, where drop casting in combination with stamping of a patterned elastomer mold yielded clusters of nanoparticles.[35] Capillarity-assisted particle assembly (CAPA) is another variation that allows a high degree of control of the number of particles per unit cell, down to single particle precision.[35,36] In earlier work, we have shown that this method can produce nanoparticle assemblies on a large scale ($>1$ cm$^2$) and can feature high-quality SLRs.[26,37]

In this study, we exploit the template-assisted approach to fabricate nanolasers using single, monocrystalline Ag nanocubes arranged in a subwavelength lattice (Fig. 1). By leveraging CAPA we successfully formed square plasmonic arrays featuring a sharp SLR. We then demonstrate lasing in the presence of a fluorescent dye with a threshold comparable to that of substrates developed using lithography-based methods. We comprehensively characterize the lasing emission, including

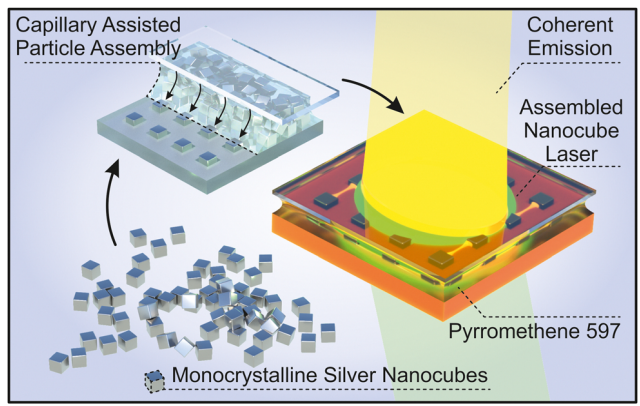

Fig. 1 A nanolaser based on assembled colloidal nanoparticles. Schematic of an optically pumped nanolaser device comprising a self-assembled array of monocrystalline Ag nanocubes and organic dye. When pumped beyond the threshold using 515 nm light, the device lases at 574 nm, consistent with the fluorescence of the pyrromethene-597 dye and surface lattice resonance-based feedback.

parameters such as emission spectra, threshold, beam profile, and polarization.

## Results

### SLR-based feedback design and nanolaser fabrication

**Nanolaser design.** The design of a nanolaser with SLR-based feedback, not unlike any other optically pumped laser, requires a good match between material properties and excitation conditions. Achieving population inversion and stimulated emission at the designed wavelength relies on engineering a spectral overlap of the three major components: the pump needs to coincide sufficiently with the absorption of the gain medium, and the photoluminescence (PL) of the gain needs to match the resonant frequency of the resonator. We chose the pumping source to be our entry point to this problem and opted to use the second harmonic of the Yb:KGW laser at 515 nm, emitting 270 fs pulses. Pyrromethene 597[38,39] (P597) was then an excellent spectral match for the pump wavelength with a high conversion rate ($>30\%$) and a proven track record as a laser dye.[40,41] P597 has a strong emission peak centered at $\sim$580 nm, which is attainable for silver photonics.

The next step is designing the feedback mechanism, which in this case is the SLR of the plasmonic Ag nanocube array. The empty lattice band structure (Fig. 2A) of a square array with a periodicity of 400 nm in the PDMS environment ($n = 1.42$) features a flat band (zero group velocity) at the $\Gamma$ point with a wavelength of $\sim$562 nm, matching the PL of the dye (Fig. 2B). A high density of optical states is thus expected at this point, which may be sufficient to provide lasing feedback.[7] We then populate this lattice using plasmonic nanoparticles to form a high-quality SLR.

The quality factor of the SLR is highest when the localized surface plasmon resonance (LSPR) of individual nanoparticles is somewhat blue-shifted with respect to the Rayleigh anomaly (RA) of the lattice (marked as solid lines in Fig. 2A and dashed lines in Fig. 2C and D). Since we have established the spectral location of the flat band in the previous step, the target is to find nanoparticles with a blue-shifted scattering peak and minimal absorption. We use Ag nanocubes because they can be synthesized in a monocrystalline form with smooth facets and we expect them to have an improved scattering-to-absorbance ratio compared to vacuum-deposited material.[42] We ran electromagnetic calculations using a nanocube geometry and found that $a$ = 65 nm edge length particles will fit the scattering requirement (blue shift with respect to the RA) very well (Fig. 2C shows the calculated scattering cross-section). In subsequent calculations, we set periodic boundary conditions with a periodicity of 400 nm. Unsurprisingly, the system features a strong resonance at 581 nm (Fig. 2D), and the electric field intensity distribution at this wavelength (Fig. 2E) showed a standing wave pattern as well as strong hot spots around the nanoparticles, characteristic of the delocalized photonic-plasmonic SLR mode. The calculated dip in transmittance has a FWHM of ∼5.8 nm, resulting in a theoretical $Q$ factor of ∼100 and overlapping the PL of the gain.

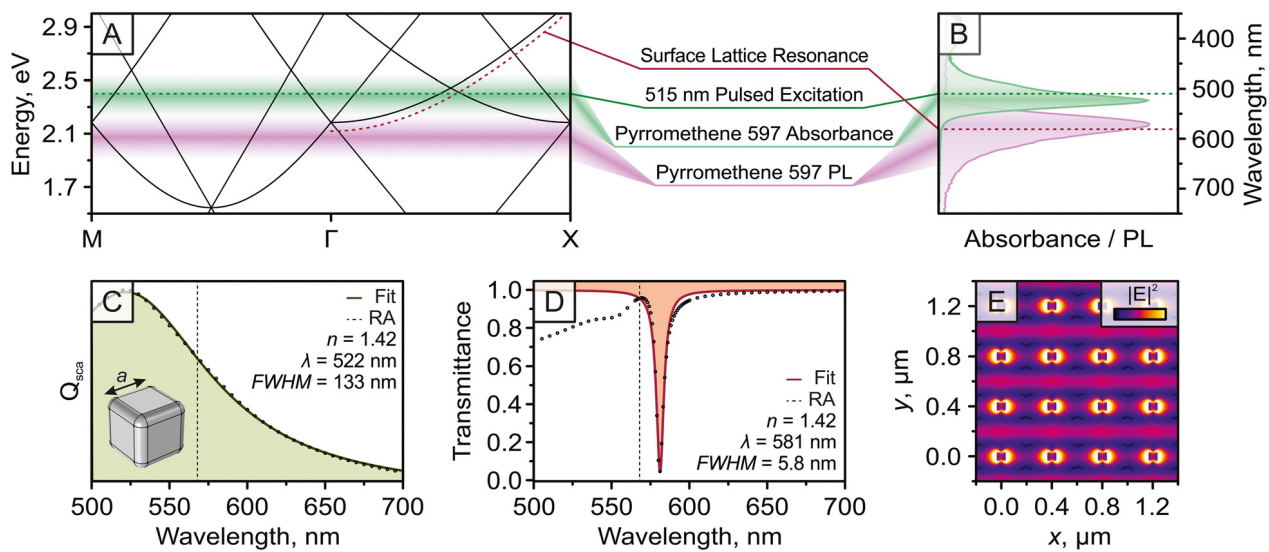

**Fig. 2** Nanolaser design. (A) Empty square lattice band structure ($\Lambda$ = 400 nm, $n$ = 1.42) with an indicated SLR location (dashed red line) forming a flat band at the $\Gamma$ point. Nanolaser pumping wavelength is indicated by a dashed green line, overlapping the absorbance of the dye. (B) The photoluminescence and absorbance of the gain (P597 laser dye). (C) Calculated scattering cross-section spectrum of $a$ = 65 nm edge length Ag nanocubes. The vertical dashed line indicates the location of the RA at the $\Gamma$ point. (D) Calculated transmittance spectrum of Ag nanocubes in the array (shaded curve is a Lorentzian fit and the location of the Rayleigh anomaly is indicated by a vertical dashed line). (E) Calculated electric field intensity distribution of Ag nanocubes in the array at the SLR wavelength ($\lambda$ = 581 nm) when illuminated by a normally incident, $x$-polarized plane wave.

**Nanolaser fabrication.** We used CAPA to position the synthesized colloidal nanoparticles with single-particle resolution (see the ESI† for the synthesis method).[43] To use this technology, a patterned template is required, which we produce by molding a PDMS replica from a silicon master prepared by common nanofabrication routes (Fig. 3A). A distinct advantage is that the replication process can be repeated multiple times without compromising the structure, minimizing the need for expensive lithography processes associated with nanofabrication while ensuring reproducibility (Fig. 3B). A scheme of the CAPA deposition process is shown in Fig. 3C, where the colloidal solution is injected between the PDMS template and a stationary glass slide. The template is then translated relative to the latter, and its temperature is elevated until nanoparticles accumulate at the meniscus. The nanocubes are coated by a thin layer of poly-vinyl-pyrrolidone (PVP, ∼2 nm). This aids the stability of the colloid, especially since dimethylformamide that we use is a theta-solvent for PVP. Colloidal stability is of utmost importance for single particle assembly, as agglomerating nanoparticles can either get stuck in the colloidal solution and prevent single particles to access the template, or get deposited on the template as a large defect, which also distorts the assembly process around it. Under optimal conditions, the nanoparticles are trapped in the pattern on PDMS with a single-particle resolution as the solution withdraws. An SEM micrograph of an assembly is shown in Fig. 3D and a full sample camera photograph, optical dark-field micrograph, and a large area SEM micrograph are displayed in the ESI,† Fig. S1.

There is some unintended disorder within these arrays, similar to the purposely introduced disorder reported in previous work.[25] The measured extinction of the assembly (Fig. 3E, red curve) still showed a high-quality SLR resonance ($\lambda$ = 572 nm, $Q$ = 66), which matches the calculation in Fig. 2D and overlaps the gain PL (Fig. 3E, pink curve).

After assembly, the substrates are preemptively soaked in a 5 mM P597 solution in DMSO:EtOH (2:1) overnight, improving stability and preventing dye diffusion into PDMS during experiments (Fig. 3C). Finally, a drop of dye solution is placed on the substrate and covered with a cover slip. We found that the nanoparticles stay in place because the trap size is very close to the dimensions of the particles, leaving no room for movement and even requiring substantial persuasion to remove them.

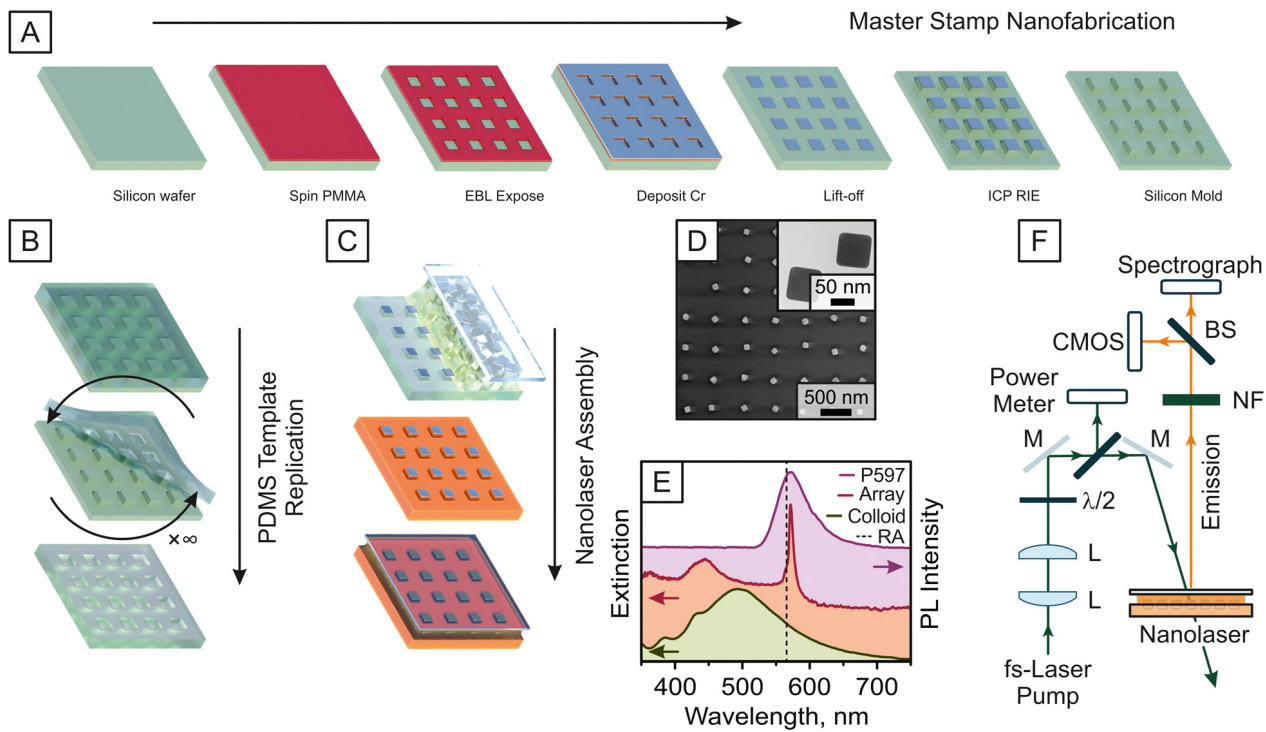

Fig. 3 Nanolaser fabrication based on nanoparticle self-assembly. (A) A master stamp is designed for a lattice resonance at a target wavelength and fabricated following standard lithography patterning and etching routines. (B) The pattern is transferred to a PDMS film via soft-lithography and this process can be repeated many times, providing templates for numerous devices. (C) Ag nanocubes featuring a LSPR close to the target wavelength are assembled into the template using CAPA. The devices are then soaked in a dye solution before being interfaced with an additional drop on top, confined by a glass cover slip. (D) An SEM micrograph of assembled nanocubes in a PDMS template. Inset shows a TEM micrograph of the nanocubes. (E) Experimental spectra of nanocubes in solution vs. in array, and the PL spectrum of the dye. The vertical dashed line indicates the location of the RA at the Γ point in PDMS. (F) The nanolasers are characterized after pumping with a 270 fs 515 nm laser using a spectrograph and a CMOS camera (L – beam reducer; $\lambda/2$ – half-waveplate, M – mirrors, NF – neutral density and notch filters).

Fig. 3F shows the experimental setup to evaluate nanolasing characteristics. We pump our devices with a collimated, Yb:KGW second harmonic 515 nm wavelength, 2.4 mm $(1/e^2)$ size beam spot with a controlled linear polarization state. We used 50 Hz frequency to avoid excessive dye bleaching and collect normal emission from the sample in free space using a fiber-coupled spectrograph and a CMOS camera.

**Nanolaser characterization**

Claims of lasing typically require three key features indicating the required level of spectral and temporal coherence of the emitted light: (i) a clear threshold of pumping power beyond which the slope of the input–output curve changes; (ii) a significant narrowing of the PL peak as the system crosses the lasing threshold, indicating the transition from spontaneous to stimulated emission; (iii) the emitted light must form a beam typical of the resonator.[44] Fig. 4 summarizes our results regarding each of these points.

**Threshold and linewidth.** We ran all characterization experiments using two pump polarizations: TM and TE, schematically shown in Fig. 4A and E, respectively. The characterization of lasing threshold is shown in Fig. 4B and F. As the pump fluence increased above the threshold value, the intensity of the emission increased faster than that of the PL emission background, which is attributed to the nonlinearity associated with stimulated emission. The PL intensity before the threshold is very small in our data because of how the experiment is set up – the spectrograph is coupled with a low-NA fiber coupler and located far from the sample, effectively eliminating all highly divergent PL signals. We recorded a threshold of 80 μJ cm$^{-2}$ using both polarizations by repeatedly scanning the fluence close to the threshold to acquire more data points. These results could be further improved by precision engineering the overlap between the SLR and the gain emission, as well as experimenting with pumping strategies.

We believe we achieve this competitive, and in many cases superior, threshold compared to reports of similar systems[11,21,29,35,45,46] (see the ESI,† Table S1, where we summarize the characteristics of similar systems in the literature), despite the unideal nanoparticle lattice because of enhanced plasmonic nanoparticle qualities enabled by monocrystallinity and low facet roughness, which decrease if not eliminate electron scattering at crystal boundaries and associated absorptive damping.[28] This experimental observation positions our work as a significant step forward towards scalable alternative to standard lithography based solutions, connecting the dots between SLR-based light-emitting nanophotonics, nanoparticle statistics, and colloid based template-assisted nanoparticle assembly.

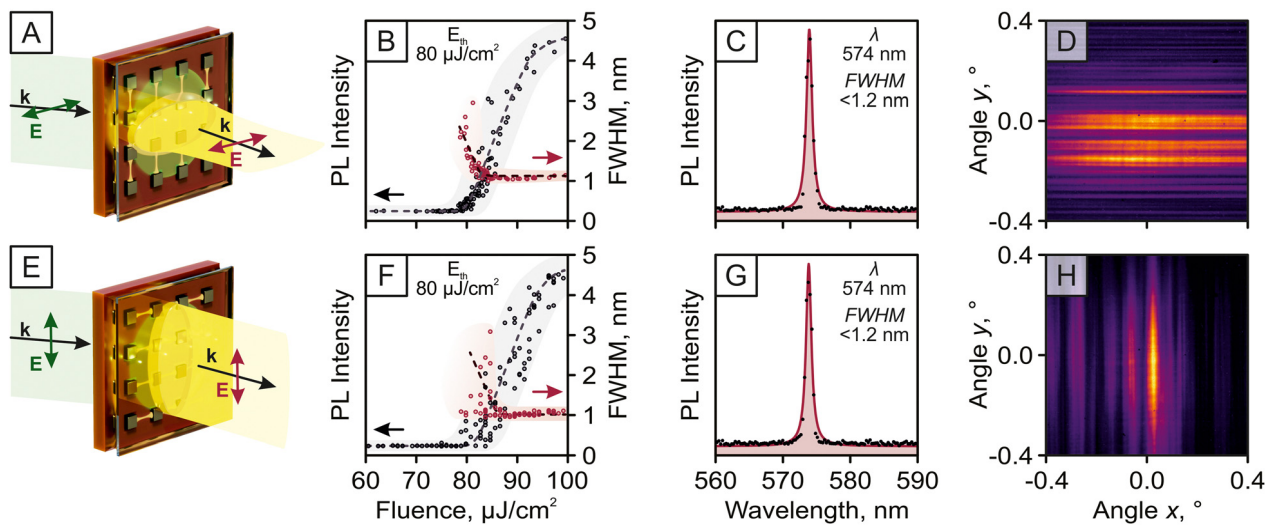

Fig. 4 Nanolaser characteristics. (A) and (E) The nanolaser emits a beam that follows the polarization of the pump and is extended in the same direction. (B) and (F) The dependence of emission intensity and the FWHM of the lasing peak on the pumping fluence, exhibiting a clear threshold behavior. (C) and (G) Emission spectra collected normal to the nanolaser plane above the lasing threshold. (D) and (H) The beam profile of the nanolaser in angular coordinates captured by a CMOS camera 22.5 cm away from the device with no collection optics.

The rapid increase in intensity beyond the threshold is accompanied by the emergence of a narrow peak around the SLR wavelength, suggesting the onset of lasing (Fig. 4C and G). Lasing occurred close to the flat band edge ($\sim$574 nm) of the lattice as intended (Fig. 2A). We observed slight wavelength variations because of the open cavity system, leading to instability of the refractive index environment during the experiment, i.e., evaporation and diffusion of the dye solvent. The PL linewidth starts at $\sim$45 nm FWHM (Fig. 2B) and rapidly decreases above the threshold to $<$1.2 nm, limited by our spectrograph resolution. In some cases, we observe a slight increase in linewidth with pumping power, but this is not uncommon for similar nanolasers.[45] Spectral narrowing is a telltale sign of temporal coherence, although a better evaluation is using the second-order correlation function $g^{(2)}$. Unfortunately, this measurement was not available at this time, and even if it was, it is challenging to acquire accurate results for a plasmonic nanolaser with gain materials composed of organic dye.[7]

**Beam profile and polarization.** The images of the laser beam were taken above the threshold by placing a CMOS camera at varying distances. Fig. 4D and H show these images taken at 22.5 cm for the two orthogonal polarizations, converted to angular coordinates (images at different distances are shown in Fig. S2, ESI†). In both cases, the beam consists of a multitude of horizontal or vertical lines, indicating that the device operates similarly to 1D distributed feedback (DFB) lasers.[30,47] We believe that the number and density of lines are associated with the quality of the nanoparticle lattice. We measured $1/e^2$ of the narrow dimension of one of the lines (which remains almost constant at varying distances) profile in angular coordinates and estimated the divergence of individual lines to $<$1 mrad.

The output beam of the nanolaser above the threshold was strongly linearly polarized and followed one of the square lattice directions. If the polarization of the exciting beam did not match the lattice orientation, the emission still did, with variations in intensity corresponding to the projection of the electric field vector of the pump onto the lattice direction. A characterization of the polarization state is displayed in Fig. S3 (ESI†).

## Discussion

Apart from the characteristics discussed above, other phenomena similar to lasing must be ruled out.[44,48,49] Specifically, amplified spontaneous emission (ASE) can appear similar to lasing. It involves the stimulated amplification of PL by a single pass through the gain volume, irrespective of the feedback mechanism.[50] This effect can especially be difficult to discriminate against lasing if the designed wavelength is very close to the PL peak of the emission, such as in our case (Fig. 2B). Like lasing, ASE can show a threshold behavior (albeit weaker and less consistent), linewidth narrowing, and produce somewhat directional emission. But the FWHM is usually in the range of a couple of tens of nm, the emitted wavelength does not depend on resonator configuration and the emission is unpolarized, contrary to the data presented here.

In our results, the FWHM is consistently $<$1.2 nm, which is the resolution limit of our spectrograph. We also tested our setup using a slightly different resonator (420 nm lattice, same nanocubes). The lasing peak followed this change (Fig. S4, ESI†), indicating that emission strongly depends on the resonator. Moreover, the beam our devices emit is clearly linearly polarized and follows the resonator geometry. Finally, in some cases we could observe the emergence of ASE at higher pumping powers, when it starts to outcompete lasing (Fig. S5, ESI†). This observation allows us to unambiguously distinguish between these emission modes.

Another process that may be mistaken for lasing is the edge-emission of waveguided modes. In this case, the threshold behavior is unexpected, which we consistently observe in our system. Moreover, we collect our data at the normal to the device surface. 1D distributed feedback (DFB) lasers emit elongated beams with low divergence along the feedback axis and high divergence perpendicular to it, which emerges as a fan-like beam.[44,51] We observe similar fan-like emission with orientation aligned to one of the lattice directions. This observation strongly suggests that the 2D array of nanoparticles essentially behaves as a collection of 1D arrays and emits a set of these fan-like beams. Plasmonic nanoparticles predominantly support electric modes, and dipolar electric modes mainly radiate, couple, and provide feedback in directions orthogonal to the axis of the induced dipole. 2D feedback and annular beams can be achieved by simultaneously providing TM and TE excitation by switching to waveguide-hybridized SLR modes (nanoparticles on a thin film)[16] or high-NA objective-based excitation.[52]

Finally, we left our devices operating and measured the expected lifetime (Fig. S6, ESI†). Although we were using an open system and a liquid dye solution, these nanolasers still maintained narrow linewidth emission for 15–30 minutes, depending on polarization. TE configuration extended the lifetime approx. 2-fold, but at a lower overall intensity. Replacement of the dye solution allows the used devices to regenerate and they can be operated again, indicating that the resonator is not damaged and the nanoparticles stay in place. Nanoparticle assemblies themselves and the resulting SLR stay stable for a long time.[26] Unless the nanoparticles are locally affected by pumping radiation—either changing shape and consequently losing the SLR, or getting dislodged from the template, the device can be operated indefinitely, as long as the dye solution is stable as indicated above.

The performance of our devices could be further improved: the threshold could be lowered by fine-tuning the interplay between the dye emission and the SLR, by setting up more efficient pumping schemes and conditions, as well as enhancing the experimental methods to more accurately estimate the onset of lasing; the efficiency could be further improved by selecting more efficient fluorescent materials, such as quantum dots as well as envisioning more efficient pumping and device schemes to suppress the loss channels such as amplified spontaneous emission, edge emission, waveguided in-plane emission, *etc.*

Overall, the demonstrated results clearly show that the emission from our devices strongly corresponds to the resonator used with a clearly defined threshold and polarization state. Therefore, our experimental observations strongly favor the claim of nanolasing from these assembled colloidal nanocube arrays. We believe our results tie together the three important facets of light-emitting nanophotonics devices: SLR-based coherent light emission, large scale colloid-based template-assisted nanoparticle assembly, and nanoparticle statistics indicating the small relevance of disorder in these systems. We showed experimentally, that such devices can be produced on a $cm^2$ scale, maintaining good light emission characteristics that are comparable or in some cases superior to lithography-based polycrystalline counterparts. We expect important applications using such platforms to follow in areas like sensing, on-chip communications, integration with flexible substrates, and structured light,[53,54] especially where our technology can be leveraged for large scale production.

## Conclusions

We presented a novel approach to develop a surface lattice resonance-based plasmonic nanolaser with competitive characteristics. We used a template-assisted technique to assemble highly crystalline silver colloidal nanocubes into periodic arrays. These plasmonic substrates featured a high-quality SLR, which served as a feedback mechanism for stimulated emission. We demonstrated a nanolaser with a narrow linewidth of $<1.2$ nm, a lasing threshold of $<100$ μJ $cm^{-2}$, and a beam characteristic of the resonator with minimal divergence and a clear polarization state. We claim that our devices are comparable to lithography-made counterparts by exploiting the enhanced emission by defects in the lattice while maintaining a low threshold and scalability. These attributes, coupled with the potential for large-scale production leveraging colloidal methods and template-assisted assembly techniques, position our devices as promising candidates for applications. Such plasmonic arrays can be inexpensively transferred onto almost any substrate, allowing easy integration in applications ranging from point-of-care diagnostics to photonic integrated circuits and optical communications systems.

## Author contributions

Conceptualization: MJ, JH, TT; data curation: MJ, NK; formal analysis: MJ, NK; funding acquisition: ST; investigation: MJ, NK, GK; methodology: MJ, TT; project administration: MJ, TT, ST; resources: JH, TT, ST; software: MJ; supervision: MJ, TT, ST; validation: MJ, NK, GK; visualization: MJ; writing – original draft: MJ, NK, GK; writing – review & editing: MJ, NK, GK, JH, TT, ST.

## Data availability

Data supporting the findings of this study are available at **https://doi.org/10.5281/zenodo.13890740** and from the corresponding author upon reasonable request.

## Conflicts of interest

There are no conflicts to declare.

## Acknowledgements

This research was performed within project "LaSensA" under the M-ERA.NET scheme and was funded by the Research Council of Lithuania (LMTLT), agreement No. S-M-ERA.NET-21-2, National Science Centre (Poland), agreement No. UMO-2020/02/Y/ST5/00086, Saxon State Ministry for Science, Culture and Tourism (Germany) and co-financed with tax funds on the basis of the budget passed by the Saxon state parliament. JH thanks the Japan Society for the Promotion of Science (JSPS) Grants-in-Aid for Scientific Research Kakenhi Program (20K05453).

## References

1 Y. Liang, C. Li, Y.-Z. Huang and Q. Zhang, Plasmonic Nanolasers in On-Chip Light Sources: Prospects and Challenges, *ACS Nano*, 2020, **14**, 14375–14390.
2 R.-M. Ma and R. F. Oulton, Applications of nanolasers, *Nat. Nanotechnol.*, 2019, **14**, 12–22.
3 N. Meinzer, W. L. Barnes and I. R. Hooper, Plasmonic meta-atoms and metasurfaces, *Nat. Photonics*, 2014, **8**, 889–898.
4 A. D. Utyushev, V. I. Zakomirnyi and I. L. Rasskazov, Collective lattice resonances: Plasmonics and beyond, *Rev. Phys.*, 2021, **6**, 100051.
5 J. Guan, *et al.*, Light–Matter Interactions in Hybrid Material Metasurfaces, *Chem. Rev.*, 2022, **122**, 15177–15203.
6 A. Vaskin, R. Kolkowski, A. F. Koenderink and I. Staude, Light-emitting metasurfaces, *Nanophotonics*, 2019, **8**, 1151–1198.
7 W. Zhou, *et al.*, Lasing action in strongly coupled plasmonic nanocavity arrays, *Nat. Nanotechnol.*, 2013, **8**, 506–511.
8 A. H. Schokker and A. F. Koenderink, Lasing at the band edges of plasmonic lattices, *Phys. Rev. B: Condens. Matter Mater. Phys.*, 2014, **90**, 155452.
9 A. Yang, *et al.*, Real-time tunable lasing from plasmonic nanocavity arrays, *Nat. Commun.*, 2015, **6**, 6939.
10 D. Wang, *et al.*, Stretchable Nanolasing from Hybrid Quadrupole Plasmons, *Nano Lett.*, 2018, **18**, 4549–4555.
11 R. Heilmann, K. Arjas and T. K. Hakala, & Päivi Törmä. Multimode Lasing in Supercell Plasmonic Nanoparticle Arrays, *ACS Photonics*, 2023, **10**, 3955–3962.
12 F. Freire-Fernández, *et al.*, Quasi-Random Multimetallic Nanoparticle Arrays, *ACS Nano*, 2023, **17**, 21905–21911.
13 T. K. Hakala, *et al.*, Lasing in dark and bright modes of a finite-sized plasmonic lattice, *Nat. Commun.*, 2017, **8**, 13687.
14 F. Freire-Fernández, *et al.*, Magnetic on–off switching of a plasmonic laser, *Nat. Photonics*, 2022, **16**, 27–32.
15 J. Guan, *et al.*, Engineering Directionality in Quantum Dot Shell Lasing Using Plasmonic Lattices, *Nano Lett.*, 2020, **20**, 1468–1474.
16 J. Guan, *et al.*, Quantum Dot-Plasmon Lasing with Controlled Polarization Patterns, *ACS Nano*, 2020, **14**, 3426–3433.
17 R. K. Yadav, *et al.*, Room Temperature Weak-to-Strong Coupling and the Emergence of Collective Emission from Quantum Dots Coupled to Plasmonic Arrays, *ACS Nano*, 2020, **14**, 7347–7357.
18 D. Xing, *et al.*, Ligand Engineering and Recrystallization of Perovskite Quantum-Dot Thin Film for Low-Threshold Plasmonic Lattice Laser, *Small*, 2022, **18**, 2204070.
19 A. Fernandez-Bravo, *et al.*, Ultralow-threshold, continuous-wave upconverting lasing from subwavelength plasmons, *Nat. Mater.*, 2019, **18**, 1172–1176.
20 J. Guan, *et al.*, Plasmonic Nanoparticle Lattice Devices for White-Light Lasing, *Adv. Mater.*, 2023, **35**, 2103262.
21 J. Guan, *et al.*, Far-field coupling between moiré photonic lattices, *Nat. Nanotechnol.*, 2023, 1–7.
22 W. Wang, *et al.*, Ultrafast Dynamics of Lattice Plasmon Lasers, *J. Phys. Chem. Lett.*, 2019, **10**, 3301–3306.
23 J. Henzie, M. H. Lee and T. W. Odom, Multiscale patterning of plasmonic metamaterials, *Nat. Nanotechnol.*, 2007, **2**, 549–554.
24 A. H. Schokker and A. F. Koenderink, Lasing in quasi-periodic and aperiodic plasmon lattices, *Optica*, 2016, **3**, 686–693.
25 A. H. Schokker and A. F. Koenderink, Statistics of Randomized Plasmonic Lattice Lasers, *ACS Photonics*, 2015, **2**, 1289–1297.
26 M. Juodėnas, T. Tamulevičius, J. Henzie, D. Erts and S. Tamulevičius, Surface Lattice Resonances in Self-Assembled Arrays of Monodisperse Ag Cuboctahedra, *ACS Nano*, 2019, **13**, 9038–9047.
27 A. R. Tao, S. Habas and P. Yang, Shape Control of Colloidal Metal Nanocrystals, *Small*, 2008, **4**, 310–325.
28 N. Khinevich, *et al.*, Size and crystallinity effect on the ultrafast optical response of chemically synthesized silver nanoparticles, *J. Materiomics*, 2024, **10**, 594–600.
29 S. Deng, *et al.*, Ultranarrow plasmon resonances from annealed nanoparticle lattices, *Proc. Natl. Acad. Sci.*, 2020, **117**, 23380–23384.
30 M. P. Knudson, *et al.*, Polarization-Dependent Lasing Behavior from Low-Symmetry Nanocavity Arrays, *ACS Nano*, 2019, **13**, 7435–7441.
31 V. Karanikolas, *et al.*, Plasmon-Triggered Ultrafast Operation of Color Centers in Hexagonal Boron Nitride Layers, *ACS Omega*, 2023, **8**, 14641–14647.
32 Y. Yin, Y. Lu, B. Gates and Y. Xia, Template-Assisted Self-Assembly: A Practical Route to Complex Aggregates of Monodispersed Colloids with Well-Defined Sizes, Shapes, and Structures, *J. Am. Chem. Soc.*, 2001, **123**, 8718–8729.
33 Y. Cui, *et al.*, Integration of Colloidal Nanocrystals into Lithographically Patterned Devices, *Nano Lett.*, 2004, **4**, 1093–1098.
34 L. Malaquin, T. Kraus, H. Schmid, E. Delamarche and H. Wolf, Controlled Particle Placement through Convective and Capillary Assembly, *Langmuir*, 2007, **23**, 11513–11521.

35 Y. Conti, N. Passarelli, J. Mendoza-Carreño, L. Scarabelli and A. Mihi, Colloidal Silver Nanoparticle Plasmonic Arrays for Versatile Lasing Architectures via Template-Assisted Self-Assembly, *Adv. Opt. Mater.*, 2023, **11**, 2300983.

36 J. Henzie, S. C. Andrews, X. Y. Ling, Z. Li and P. Yang, Oriented assembly of polyhedral plasmonic nanoparticle clusters, *Proc. Natl. Acad. Sci.*, 2013, **110**, 6640–6645.

37 M. Juodėnas, *et al.*, Effect of Ag Nanocube Optomechanical Modes on Plasmonic Surface Lattice Resonances, *ACS Photonics*, 2020, **7**, 3130–3140.

38 D. M. AL-Aqmar, H. I. Abdelkader and M. T. H. Abou Kana, Optical, photo-physical properties and photostability of pyrromethene (PM-597) in ionic liquids as benign green-solvents, *J. Lumin.*, 2015, **161**, 221–228.

39 W. P. Partridge, N. M. Laurendeau, C. C. Johnson and R. N. Steppel, Performance of Pyrromethene 580 and 597 in a commercial Nd:YAG-pumped dye-laser system, *Opt. Lett.*, 1994, **19**, 1630–1632.

40 Z. Hu, *et al.*, Coherent Random Fiber Laser Based on Nanoparticles Scattering in the Extremely Weakly Scattering Regime, *Phys. Rev. Lett.*, 2012, **109**, 253901.

41 E. Yariv, S. Schultheiss, T. Saraidarov and R. Reisfeld, Efficiency and photostability of dye-doped solid-state lasers in different hosts, *Opt. Mater.*, 2001, **16**, 29–38.

42 D. Peckus, *et al.*, Hot Electron Emission Can Lead to Damping of Optomechanical Modes in Core–Shell Ag@TiO2 Nanocubes, *J. Phys. Chem. C*, 2017, **121**, 24159–24167.

43 T. Kraus, *et al.*, Nanoparticle printing with single-particle resolution, *Nat. Nanotechnol.*, 2007, **2**, 570–576.

44 I. D. W. Samuel, E. B. Namdas and G. A. Turnbull, How to recognize lasing, *Nat. Photonics*, 2009, **3**, 546–549.

45 H. T. Rekola, T. K. Hakala and P. Törmä, One-Dimensional Plasmonic Nanoparticle Chain Lasers, *ACS Photonics*, 2018, **5**, 1822–1826.

46 T. B. Hoang, G. M. Akselrod, A. Yang, T. W. Odom and M. H. Mikkelsen, Millimeter-Scale Spatial Coherence from a Plasmon Laser, *Nano Lett.*, 2017, **17**, 6690–6695.

47 K. Yoshida, *et al.*, Electrically driven organic laser using integrated OLED pumping, *Nature*, 2023, **621**, 746–752.

48 Scrutinizing lasers, *Nat. Photonics*, 2017, **11**, 139.

49 R. Sato, *et al.*, Random Lasing via Plasmon-Induced Cavitation of Microbubbles, *Nano Lett.*, 2021, **21**, 6064–6070.

50 A. J. C. Kuehne and M. C. Gather, Organic Lasers: Recent Developments on Materials, Device Geometries, and Fabrication Techniques, *Chem. Rev.*, 2016, **116**, 12823–12864.

51 I. D. W. Samuel and G. A. Turnbull, Organic Semiconductor Lasers, *Chem. Rev.*, 2007, **107**, 1272–1295.

52 G. Heliotis, *et al.*, Emission Characteristics and Performance Comparison of Polyfluorene Lasers with One- and Two-Dimensional Distributed Feedback, *Adv. Funct. Mater.*, 2004, **14**, 91–97.

53 M. L. Fajri, *et al.*, Designer Metasurfaces via Nanocube Assembly at the Air–Water Interface, *ACS Nano*, 2024, **18**, 26088–26102.

54 J. Mendoza-Carreño, *et al.*, Nanoimprinted 2D-Chiral Perovskite Nanocrystal Metasurfaces for Circularly Polarized Photoluminescence, *Adv. Mater.*, 2023, **35**, 2210477.